\documentclass[sigconf,10pt,nonacm,9pt]{acmart} 
\usepackage{booktabs}
\usepackage{graphicx}
\usepackage{subfigure}
\usepackage{makecell}
\usepackage[ruled,vlined]{algorithm2e}
\usepackage{amsthm}

\newtheorem{remark}{Remark}

\title{Multi-Agent Reinforcement Learning for Decentralized Reservoir Management via Murmuration Intelligence}

\author{Heming Fu$^1$, Guojun Xiong$^2$, Jian Li$^1$, Shan Lin$^1$}
\affiliation{%
  \institution{$^1$Stony Brook University, $^2$Harvard University}
  \country{\{heming.fu, jian.li.3, shan.x.lin\}@stonybrook.edu, gjxiong@g.harvard.edu}
}

\renewcommand{\shortauthors}{H. Fu et al.}

\begin{document}


\begin{abstract}
  Conventional centralized water management systems face critical limitations from computational complexity and uncertainty propagation. We present \texttt{MurmuRL}, a novel decentralized framework inspired by starling murmurations intelligence, integrating bio-inspired \emph{\textbf{alignment}}, \emph{\textbf{separation}}, and \emph{\textbf{cohesion}} rules with multi-agent reinforcement learning. \texttt{MurmuRL} enables individual reservoirs to make autonomous local decisions while achieving emergent global coordination. Experiments on grid networks demonstrate that \texttt{MurmuRL} achieves 8.8\% higher final performance while using 27\% less computing overhead compared to centralized approaches. Notably, strategic diversity scales super-linearly with system size, exhibiting sophisticated coordination patterns and enhanced resilience during extreme events. \texttt{MurmuRL} offers a scalable solution for managing complex water systems by leveraging principles of natural collective behavior.
\end{abstract}

\maketitle

\section{Introduction}

\begin{figure}[t]
\centering
\subfigure[Reservoir network in Northeast Hunan, China]{
  \includegraphics[width=0.41\linewidth]{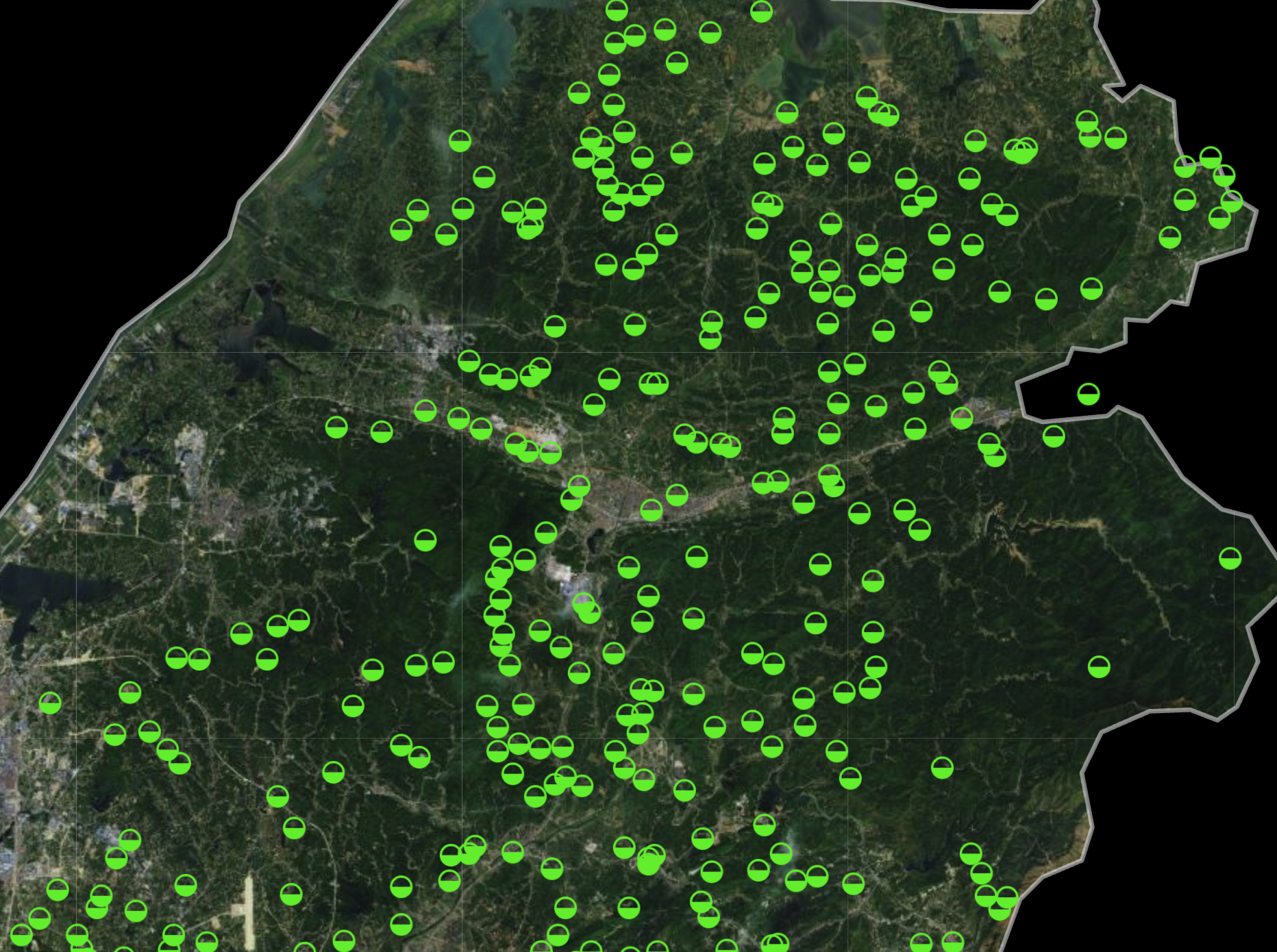}
  \label{fig:hunan_network}
}
\subfigure[Local reservoir management]{
  \includegraphics[width=0.48\linewidth]{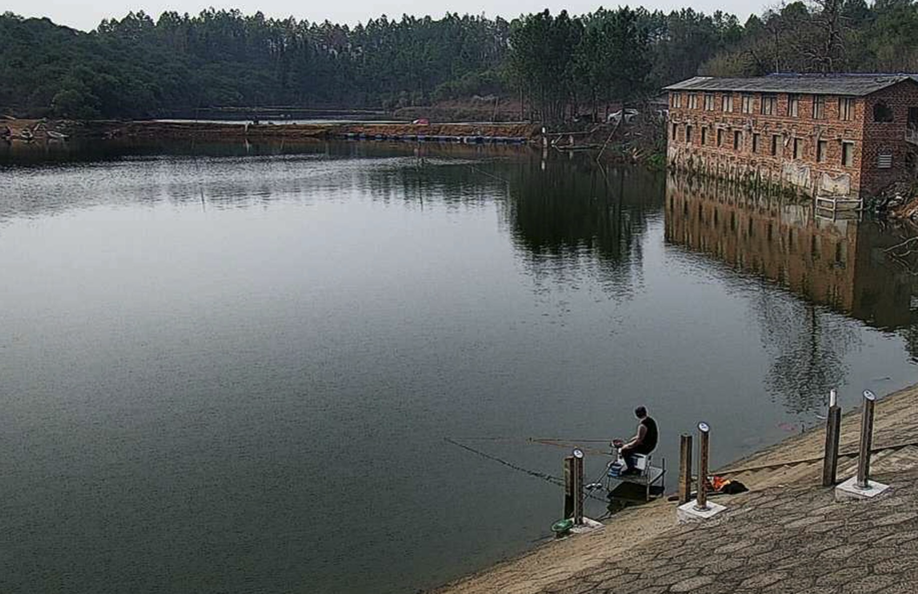}
  \label{fig:reservoir}
}
\subfigure[Global flood frequency map (1970-2018)]{
  \includegraphics[width=0.95\linewidth]{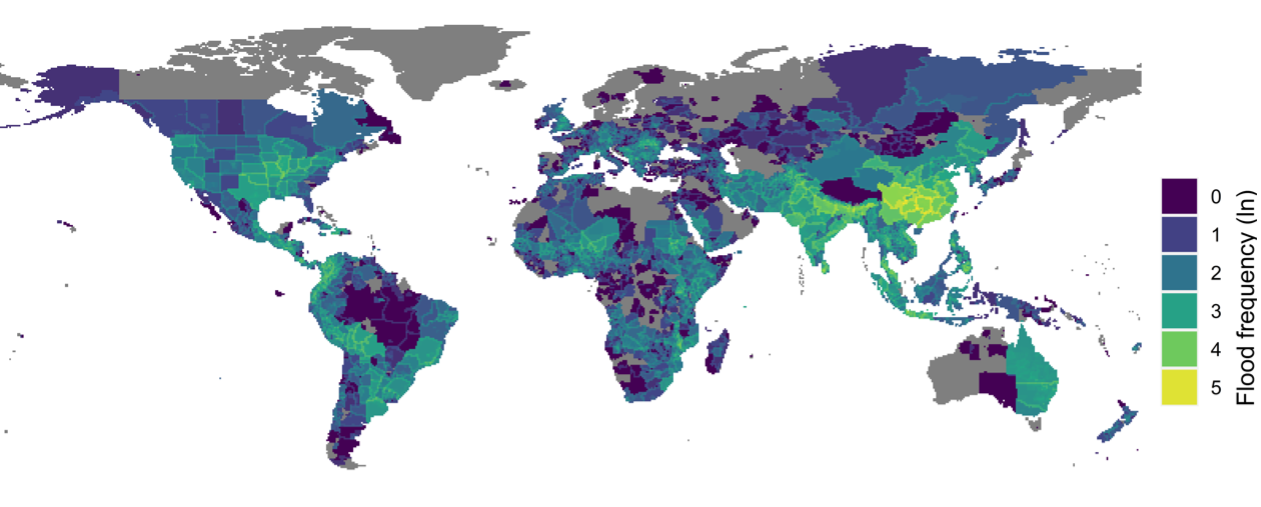}
  \label{fig:flood_map}
}
\caption{Water resource management challenges: (a) complex networks of interconnected reservoirs requiring coordination, (b) individual reservoirs that must be locally managed, and (c) high global flood frequency demanding more robust management systems~\cite{flood_displacement}.}
\label{fig:water_management}
\vspace{-0.5cm}
\end{figure}

Water resources management is critical for human welfare and environmental sustainability. Recent data highlights the severity of water-related challenges: between 2000 and 2019, flood-related disasters affected over 1.65 billion people globally, causing approximately 100,000 deaths and \$651 billion in economic losses~\cite{rentschler2022flood}. Simultaneously, drought events impacted over 1.43 billion people, with economic damages exceeding \$130 billion~\cite{unccd2022drought}. Climate change has intensified these challenges, with once-in-a-century precipitation events now occurring approximately every 20 years in many regions and flash floods increasing by 21\% in the past decade~\cite{un2022drought}.

Reservoir networks are complex interconnected water systems that serve multiple critical functions: flood prevention, water supply security, hydroelectric power generation, and ecological flow maintenance. As shown in Figure~\ref{fig:hunan_network}, modern networks like Northeast Hunan's hundreds of interconnected reservoirs create vast distributed systems spanning diverse terrains. Each individual reservoir (Figure~\ref{fig:reservoir}) requires constant management decisions that must balance competing objectives under significant uncertainty. 

For instance, maintaining high water levels optimizes drought resilience and power generation but increases flood risk during sudden precipitation events; conversely, keeping levels low mitigates flooding but compromises water security and energy production. These management challenges are exacerbated by climate change, which has intensified hydrological extremes globally (Figure~\ref{fig:flood_map}), rendering historical data patterns increasingly unreliable for prediction. Real-world reservoir networks require coordinating hundreds of daily release decisions, with each potentially affecting downstream conditions for hundreds of kilometers. Similar complex networks worldwide—including California's Central Valley Project (20 major reservoirs), Spain's Ebro River Basin (138 reservoirs), and Australia's Murray-Darling system (over 240 water storage facilities)—all face unprecedented operational challenges that conventional centralized management approaches struggle to address effectively.

Conventional centralized water management systems face two critical limitations: (1) computational complexity that grows exponentially with the number of nodes, leading to energy consumption and response latency~\cite{cheng2014complexity, labadie2004optimal}; and (2) propagation of uncertainty that renders precise global water allocation practically impossible~\cite{castelletti2010tree}. 

Existing decentralized approaches, while addressing computational challenges, often fail to achieve effective coordination across the network~\cite{zhang2021multi}. Traditional Multi-Agent Reinforcement Learning (MARL) methods typically prioritize short-term local objectives, resulting in "myopic optimization" where individual agents compete rather than cooperate~\cite{ yu2022surprising}. This leads to emergent negative behaviors such as resource hoarding during scarcity and inefficient discharge patterns during abundance, ultimately compromising system-wide resilience~\cite{castelletti2013multiobjective}.
\begin{figure}[t]
\centering
\includegraphics[width=0.45\textwidth]{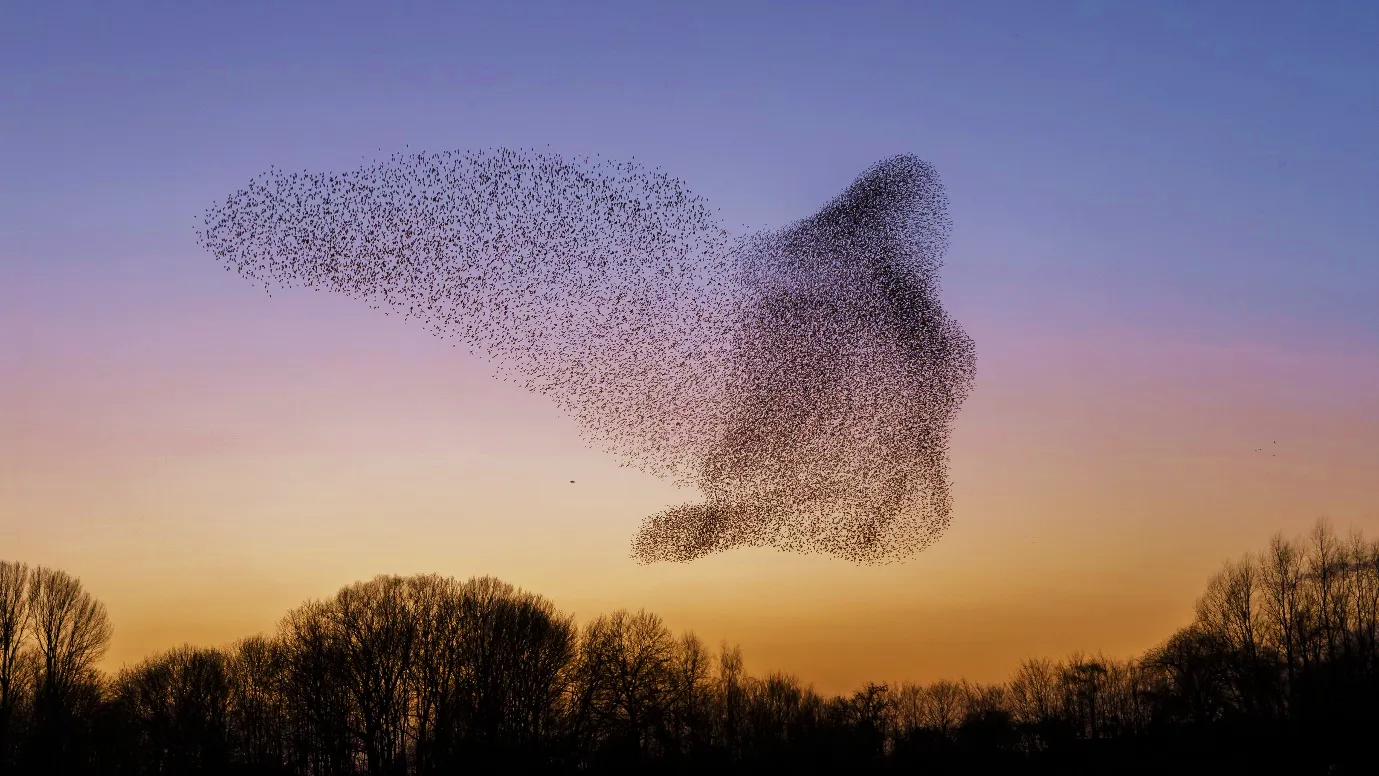}
\caption{Starling Murmuration: A natural example of emergent intelligence from simple local interaction rules, where thousands of birds create complex, responsive patterns without centralized coordination. (Photo by Menno Schaefner~\cite{schaefner_murmuration})}
\label{fig:murmuration}
\vspace{-0.5cm}
\end{figure}

Inspired by starling murmurations intelligence (Figure~\ref{fig:murmuration}), we propose a fundamentally different approach to reservoir management. Starling murmurations demonstrate how thousands of birds, each following simple local interaction rules with nearby neighbors, collectively exhibit sophisticated, responsive behavior without centralized coordination.
We introduce \texttt{MurmuRL}, a novel framework that integrates bio-inspired local interaction rules—\emph{\textbf{alignment}}, \emph{\textbf{separation}}, and \emph{\textbf{cohesion}} (ASC)—with MARL for decentralized reservoir management. The key contributions of our work are:
\begin{itemize}
\item[$\triangleright$] \textbf{Murmuration-Guided MARL Framework:} We propose the first MARL framework inspired by biological murmuration intelligence for decentralized water resource management.
\item[$\triangleright$] \textbf{Emergent Murmuration Intelligence:} Our experiments show that sophisticated global coordination emerges naturally from local interactions, with stable performance as the system size increases—a critical advantage over traditional approaches.
\item[$\triangleright$] \textbf{Energy-Efficient Performance:} \texttt{MurmuRL} achieves superior water management with up to 27\% less computational overhead compared to centralized approaches, while significantly improving response times during simulated extreme weather events.
\end{itemize}

\section{Related Work}

\subsection{Water Resource Management Systems}
Traditional water management approaches rely on optimization techniques including linear programming~\cite{yeh1985reservoir}, network flow algorithms~\cite{labadie2004optimal}, and stochastic dynamic programming for handling uncertainty~\cite{giuliani2016direct}. Recent advances have explored evolutionary algorithms~\cite{castelletti2012evolutionary} and model predictive control~\cite{schwanenberg2015forecast}. While these approaches work for systems like California's Central Valley Project (20 reservoirs) and China's Three Gorges Dam, they deteriorate significantly when scaled to larger networks or faced with increased environmental uncertainty.

\subsection{Multi-Agent Reinforcement Learning}
MARL provides frameworks for decentralized decision-making in complex control problems. Our experimental baselines include MADDPG~\cite{lowe2017multi} (centralized training with decentralized execution), QMIX~\cite{rashid2018qmix} (using mixing networks), MAPPO~\cite{yu2022surprising} (multi-agent PPO), VDN~\cite{sunehag2017value} (value decomposition), and IQL~\cite{tan1993multi} (independent learning). Water-specific MARL applications remain limited to small-scale systems~\cite{mason2018applying, castelletti2013multiobjective}, typically struggling with scalability in realistic hydrological networks.

\subsection{Starling Murmuration Theory}
Starling murmurations are a striking example of collective intelligence. Reynolds~\cite{reynolds1987flocks} distilled their behavior into three simple rules: alignment (matching neighbors' velocities), cohesion (moving toward nearby flockmates), and separation (avoiding crowding). These local interactions produce complex emergent behavior without centralized control.

Mathematical foundations were advanced by Vicsek et al.~\cite{vicsek1995novel}, who modeled phase transitions from local rules to global order, and by Cucker and Smale~\cite{cucker2007emergent}, who proved flocking convergence under various conditions. Parisi's Nobel-winning work~\cite{parisi2004critical} on spin glasses offered insights into how ordered states emerge from uncertain, decentralized systems.
These bio-inspired principles have demonstrated remarkable success across diverse domains. Particle swarm optimization~\cite{kennedy1995particle} leverages murmuration principles for complex search problems, while ant colony optimization~\cite{dorigo2006ant} applies similar concepts to telecommunications routing. Other successful applications include distributed robotic control~\cite{turgut2008self}, autonomous vehicle coordination~\cite{li2020cooperative}, and smart grid load balancing~\cite{zhang2018distributed}.

\section{Problem Formulation}

\begin{figure}[h]
\centering
\includegraphics[width=0.35\textwidth]{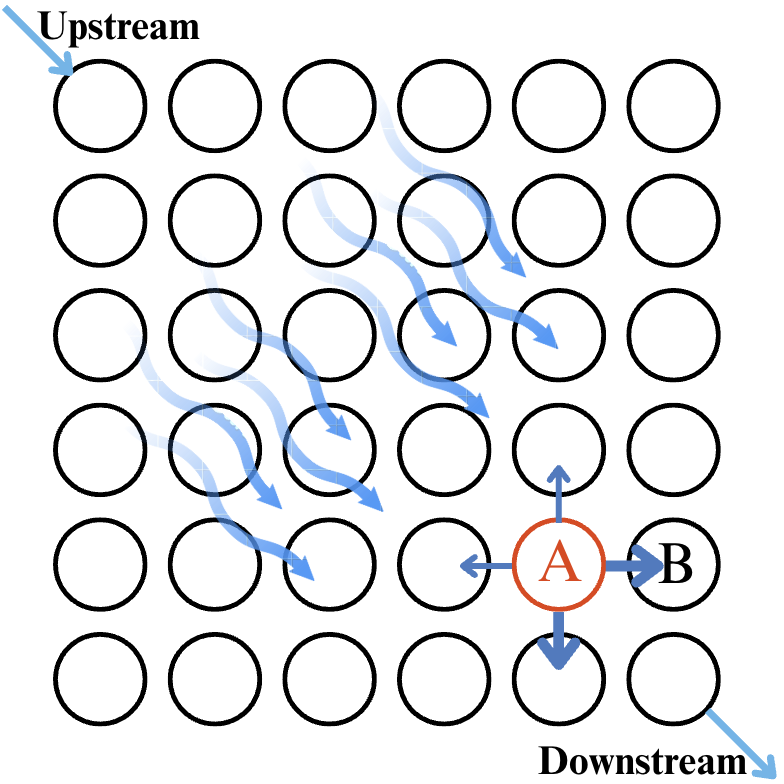}
\caption{Reservoir node modeling in a grid-based hydrological network. Each node (e.g., node A) interacts with four neighboring nodes, managing water flows in a general pattern from upstream (top-left) to downstream (bottom-right).}
\label{fig:node_model}
\end{figure}

We model the reservoir network as a stochastic graph $\mathcal{G} = (V, E)$ on an $N \times N$ grid, where $V = \{v_1, v_2, \ldots, v_{N^2}\}$ represents the set of reservoir nodes and $E$ represents water flow channels between nodes. As illustrated in Figure~\ref{fig:node_model}, each node can release water to or receive water from its four adjacent neighbors (north, east, south, west), with the overall flow pattern following a general direction from upstream (top-left) to downstream (bottom-right).

The critical challenge in this environment is that node interactions are inherently uncertain due to environmental factors. When a reservoir node $i$ attempts to release water volume $x$ to a neighboring node $j$, the actual received volume follows a stochastic distribution:
\begin{align}
f_{ij}(t) \sim \mathcal{N}(\mu_{ij}(t), \sigma_{ij}^2(t)),
\end{align}
where $\mu_{ij}(t) = x$ is the intended transfer volume, and $\sigma_{ij}^2(t)$ represents environmental uncertainty. Importantly, this uncertainty itself is not static but varies according to:
\begin{align}
\sigma_{ij}^2(t) \sim \mathcal{N}(\sigma_{\text{base}}, \delta_{ij}^2(t)),
\end{align}
where $\sigma_{\text{base}}$ is the baseline uncertainty and $\delta_{ij}^2(t)$ captures temporal and spatial variations from weather patterns, seasonal changes, and geographic factors. This dual-layer uncertainty modeling captures both variability in water transfers and in the environmental factors affecting those transfers.

Each node $i$ is characterized by its water level $h_i(t)$, inflows $q_{\text{in},i}(t)$ from external sources (precipitation), outflows $q_{\text{out},i}(t)$ to external sinks (evaporation, consumption), and control action $a_i(t)$ determining water releases. The water level dynamics follow:
\begin{align}
h_i(t+1) = h_i(t) + \frac{1}{C_i} \Big( &\sum_{j \in \mathcal{N}_{\text{in}}(i)} f_{ji}(t) - \sum_{k \in \mathcal{N}_{\text{out}}(i)} f_{ik}(t) \nonumber\\
&+ q_{\text{in},i}(t) - q_{\text{out},i}(t) \Big),
\end{align}
where $C_i$ is the capacity of reservoir $i$, $\mathcal{N}_{\text{in}}(i)$ and $\mathcal{N}_{\text{out}}(i)$ represent the sets of upstream and downstream neighbors, respectively.

This dual-layer uncertainty—both in water transfer and in the factors affecting that uncertainty—creates a cascade effect that makes traditional centralized optimization approaches computationally intractable. For a network with $n$ nodes, the uncertainty propagation means that downstream predictions after just 10 nodes can have over 40\% error, making precise global allocation practically impossible.

The objective for each reservoir $i$ is to maintain optimal water levels while contributing to system-wide flood and drought mitigation:
\begin{align}
\max_{\pi} \mathbb{E}\left[\sum_{t=0}^{T} \gamma^t \sum_{i=1}^{N^2} R_i(h_i(t), a_i(t))\right],
\end{align}

where $\pi$ represents the policy mapping states to actions, $\gamma$ is the discount factor, and $R_i$ is the reward function defined as:
\begin{align}
R_i(h_i(t), a_i(t)) = &w_{\text{target}} \cdot r_{\text{target}}(h_i(t)) + w_{\text{flood}} \cdot r_{\text{flood}}(h_i(t)) \nonumber\\
&+ w_{\text{drought}} \cdot r_{\text{drought}}(h_i(t)) \nonumber\\
&+ w_{\text{efficiency}} \cdot r_{\text{efficiency}}(a_i(t))
\end{align}
where $r_{\text{target}}$ penalizes deviation from target water levels, $r_{\text{flood}}$ penalizes excessive levels, $r_{\text{drought}}$ penalizes critically low levels, and $r_{\text{efficiency}}$ discourages wasteful releases.

\section{Methodology}

Our \texttt{MurmuRL} framework addresses the challenges of uncertainty propagation and coordination through bio-inspired local interaction rules integrated into a MARL framework.

\subsection{Bio-Inspired Murmuration Rules}

The key insight from starling murmurations is that sophisticated collective behavior can emerge from agents following three simple local rules:

\emph{\textbf{Alignment:}} In the water management context, alignment encourages neighboring reservoirs to adopt similar release strategies, creating coordinated regional policies:
\begin{align}
\mathcal{L}_{\text{align},i}(t) = \left\| a_i(t) - \frac{1}{|\mathcal{N}_i|}\sum_{j \in \mathcal{N}_i} a_j(t) \right\|^2,
\end{align}
where $\mathcal{N}_i$ represents the set of neighbors for reservoir $i$.

\emph{\textbf{Separation:}} To prevent homogenization of strategies and maintain functional diversity across the network, the separation rule ensures reservoirs maintain appropriate differentiation when needed:
\begin{align}
\mathcal{L}_{\text{sep},i}(t) = \sum_{j \in \mathcal{N}_i} \max \left( 0, 1 - \frac{\| a_i(t) - a_j(t) \|}{\tau} \right),
\end{align}
where $\tau$ is a threshold parameter. This encourages strategic diversity when conditions differ across regions.

\emph{\textbf{Cohesion:}} Promotes system-wide alignment with environmental requirements, encouraging reservoirs to collectively meet ecological flow demands:
\begin{align}  
\mathcal{L}_{\text{coh},i}(t) = \left\| \sum_{j \in \mathcal{N}_i \cup \{i\}} \phi_{ij}(t) - \phi_{\text{eco}} \right\|^2,
\end{align}
where $\phi_{\text{eco}}$ represents target ecological flows for downstream health.

\subsection{Algorithm Overview: \texttt{MurmuRL}}
We define the state for each agent $i$ to include: local water level $h_i(t)$, recent inflow/outflow history $\{f_{ji}(t-k), f_{ij}(t-k)\}_{k=0}^K$, neighbor states $\{h_j(t)\}_{j \in \mathcal{N}_i}$
and environmental forecasts $\hat{q}_{\text{in},i}(t+1:t+K)$ and $\hat{q}_{\text{out},i}(t+1:t+K)$:
\begin{align}
s_i(t) := \Big(h_i(t), \{f_{ji}(t-k), f_{ij}(t-k)\}_{k=0}^K, \{h_j(t)\}_{j \in \mathcal{N}_i}, \nonumber\\
\hat{q}_{\text{in},i}(t+1:t+K), \hat{q}_{\text{out},i}(t+1:t+K)\Big). 
\end{align}

The action $a_i(t)$ for agent $i$ consists of continuous decisions on water releases to each connected neighbor and external sinks. Each agent $i$ maintains a policy $\pi_i(a_i(t)|s_i(t))$ mapping states to actions.

The core innovation in \texttt{MurmuRL} is connecting the standard reinforcement learning objective in equation (4) with our bio-inspired coordination mechanism. While the reward function in equation (5) defines the overall objective for water management (maintaining optimal levels and preventing floods/droughts), this alone is insufficient for coordination in a decentralized setting. We integrate this with the ASC rules through an enhanced policy objective:

\begin{align}
J_i(\pi_i) = \mathbb{E}_{s,a}\Big[ R_i(s_i(t),a_i(t)) &\!-\! w_1 \mathcal{L}_{\text{align},i}(t)\!-\! w_2 \mathcal{L}_{\text{sep},i}(t)\! \nonumber\\&-\! w_3 \mathcal{L}_{\text{coh},i}(t) \Big],
\end{align}
where $R_i(s_i(t),a_i(t))$ is the individual agent's contribution to the global objective from equation (4), and $w_1$, $w_2$, and $w_3$ are weights balancing the murmuration components. This creates a direct link between the traditional RL reward (equation 5) and the bio-inspired coordination mechanism, ensuring that agents optimize both local performance and global coordination simultaneously.
Parameterizing the policy $\pi_i$ with parameters $\theta_i$, i.e., $\pi_i(a_i(t)|s_i(t);\theta_i)$ and following standard policy gradient update~\cite{sutton2000policy, silver2014deterministic}, we have:
\begin{align*}
\nabla_{\theta_i}\! J_i(\pi_i(\theta_i))\! =
\!\mathbb{E}_{s,a}\Big[ \nabla_{\theta_i}\! \log \pi_i(a_i(t)|s_i(t);\theta_i) \!\cdot\! Q_i(s_i(t),a_i(t)) \Big],
\end{align*}
where $Q_i(s_i(t),a_i(t))$ combines the reward and ASC terms. The parameter update is:
\begin{align}
\theta_i^{\tau+1} \leftarrow \theta_i^{\tau} + \alpha \nabla_{\theta_i} J_i(\pi_i(\theta_i)),
\end{align}
where $\alpha$ is the learning rate. The entire algorithm is presented in Algorithm \ref{alg:1}.
\vspace{-0.5em}
\begin{remark}
  This framework encourages emergent coordination patterns that are fundamentally different from both centralized approaches and traditional MARL~\cite{lowe2017multi, rashid2018qmix}. The alignment term promotes regional consistency, the separation term preserves strategic diversity where needed, and the cohesion term ensures system-wide objectives are met. As the system scale increases, these simple local rules generate increasingly sophisticated global coordination patterns that are particularly effective for handling the uncertainty cascade in large reservoir networks. Moreover, the computational complexity scales linearly with the number of nodes—in contrast to the exponential scaling of centralized approaches.
\end{remark}

\begin{algorithm}[t]
\caption{\texttt{MurmuRL} Framework}
\label{alg:1}
\SetAlgoLined
\SetKw{KwInit}{Initialize}
\KwInit{Policy parameters $\{\theta_i^0\}_{i=1}^n$ for all reservoir agents}

\For{each episode $\tau$}{
    Initialize environment state\;
    \For{each timestep $t$}{
        \For{each agent $i$}{
            Observe local state $s_i(t)$\;
            Execute action $a_i(t) \sim \pi_i(a_i(t)|s_i(t);\theta_i^{\tau})$\;
            Receive reward $R_i(s_i(t),a_i(t))$\;
        }
        Environment transitions to next state\;
    }
    \For{each agent $i$}{
        Compute losses: $\mathcal{L}_{\text{align},i}(t)$, $\mathcal{L}_{\text{sep},i}(t)$, $\mathcal{L}_{\text{coh},i}(t)$\;
        Compute total objective $J_i(\pi_i)$\;
        Update policy: $\theta_i^{\tau+1} \leftarrow \theta_i^{\tau} + \alpha \nabla_{\theta_i} J_i(\pi_i(\theta_i^{\tau}))$\;
    }
}
\end{algorithm}

\section{Preliminary Results}

We conducted a two-phase experimental evaluation to validate our approach, focusing first on the emergent properties of the \emph{\textbf{alignment}}, \emph{\textbf{separation}}, and \emph{\textbf{cohesion}} rules and then on the full RL framework.

\subsection{Experiment Setup}

\subsubsection{Emergent Global Intelligence at Different Scales}
To isolate and demonstrate the fundamental advantage of bio-inspired local interaction rules, we first conducted controlled experiments comparing the Murmuration-guided approach against a simple baseline at different grid scales.

We simulated both 50×50 and 100×100 node grid networks, where each node represents a reservoir connected to its four adjacent neighbors. Each node maintained a scalar action value representing its water management strategy. The simulations ran for 200 timesteps with the following parameters:

\begin{itemize}
    \item[$\triangleright$] Murmuration method: alignment coefficient $k_{align}=0.1$, separation coefficient $k_{sep}=0.1$ with threshold=1.0, cohesion coefficient $k_{coh}=0.1$;
    \item[$\triangleright$] Baseline method: local averaging coefficient $k_{base}=0.1$;
    \item[$\triangleright$] Both methods incorporated Gaussian noise with $\sigma=0.05$ to model environmental uncertainty.
\end{itemize}
We measured cluster count (number of distinct strategy clusters), local difference (average absolute difference between neighboring nodes), and temporal fluctuation (average change in node values between consecutive timesteps).

\subsubsection{\texttt{MurmuRL} Framework Evaluation}
Building on the insights from the first experiment, we implemented and evaluated the complete \texttt{MurmuRL} framework against standard MARL approaches, with details as follows:
\begin{itemize}
    \item[$\triangleright$] States included local water level, neighbor states, and inflow predictions;
    \item[$\triangleright$] Actions represented continuous water release decisions;
    \item[$\triangleright$] Reward function balanced local objectives (maintaining target water levels) and global concerns (avoiding floods/droughts);
    \item[$\triangleright$] ASC rules were incorporated with weights $w_1=0.2$, $w_2=0.1$, $w_3=0.3$.
\end{itemize}

For baselines, we used standard implementations of MADDPG, QMIX, MAPPO, VDN, and IQL with equivalent network architectures (two hidden layers of 64 units) and hyperparameters. All methods were trained for 1500 episodes with an episode length of 100 timesteps. We used Adam optimizer with a learning rate of 0.001 and discount factor $\gamma=0.99$.

\subsection{Results and Analysis}

\begin{figure}[t]
\centering
\begin{tabular}{c}
\includegraphics[width=0.45\textwidth]{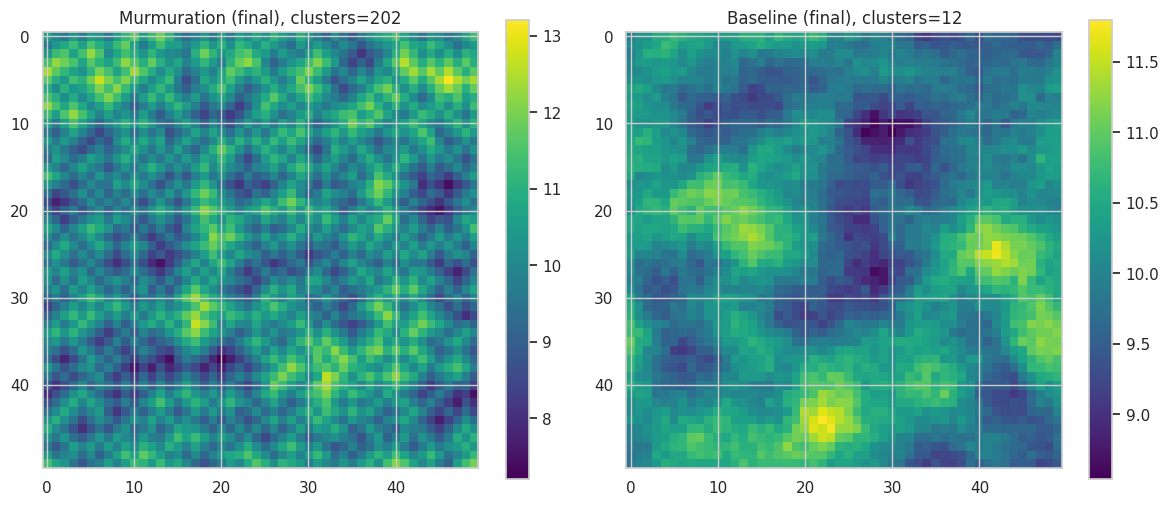} \\
\small (a) 50×50 grid: Murmuration(202 clusters) vs. Baseline(12 clusters) \\[0.3cm]
\includegraphics[width=0.45\textwidth]{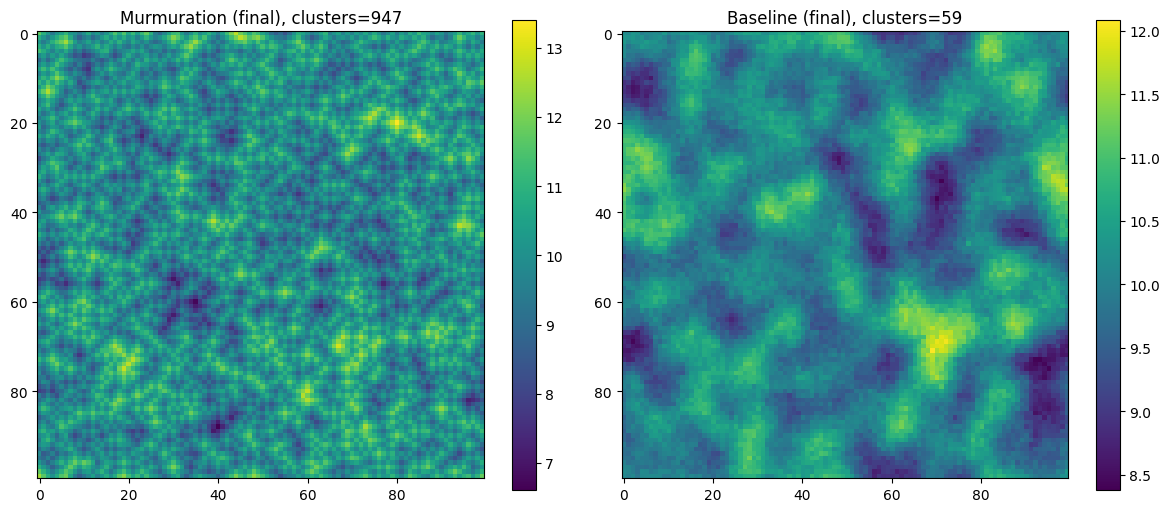} \\
\small (b) 100×100 grid: Murmuration(947 clusters) vs. Baseline(59 clusters)
\end{tabular}
\caption{Cluster formation comparison at different scales. The Murmuration approach (left images) generates significantly more distinct clusters than the baseline approach (right images) at both scales, with the advantage becoming more pronounced at the larger scale. Colors represent water management strategies with values between approximately 7 and 13 units.}
\label{fig:heatmaps}
\vspace{-0.8em}
\end{figure}

Figure~\ref{fig:heatmaps} presents visualizations of the cluster formation in water management strategies after 200 simulation timesteps at different scales. The results demonstrate a striking pattern: as the system scale increases, the advantage of our Murmuration-guided approach becomes even more pronounced.

At the 50×50 scale (Figure~\ref{fig:heatmaps}a), the Murmuration approach generates 202 distinct clusters, showing a rich pattern of strategies. In contrast, the baseline approach forms merely 12 clusters, indicating a much coarser strategy distribution. When the scale increases to 100×100 (Figure~\ref{fig:heatmaps}b), our approach produces 947 distinct clusters—a 4.7× increase despite only a 4× increase in node count—while the baseline approach only increases to 59 clusters. This demonstrates the emergence of increasingly sophisticated patterns as the system grows.

The quantitative results from both scales are summarized in Table~\ref{tab:emergence}:
\begin{table}[t]
\centering
\begin{tabular}{lcc|cc}
\toprule
\textbf{Metric} & \textbf{M} & \textbf{B} & \textbf{M} & \textbf{B} \\
& \multicolumn{2}{c|}{\textbf{50×50}} & \multicolumn{2}{c}{\textbf{100×100}} \\
\midrule
Cluster count & 202 & 12 & 947 & 59 \\
Local diff. & 0.691 & 0.160 & 0.688 & 0.161 \\
Temp. fluct. & 0.064 & 0.057 & 0.064 & 0.057 \\
\bottomrule
\end{tabular}
\caption{Comparison of emergent properties at different scales}
\label{tab:emergence}
\vspace{-2em}
\end{table}
both sets of experiments show that the proposed \texttt{MurmuRL} maintains a consistently higher local difference (approximately 0.69 vs. 0.16), indicating more diverse local strategies rather than homogenization. Importantly, this diversity comes with only a minimal increase in temporal fluctuation (0.064 vs. 0.057), demonstrating that the system maintains stability despite complex spatial patterns.

\begin{figure}[t]
  \centering
  \includegraphics[width=0.35\textwidth]{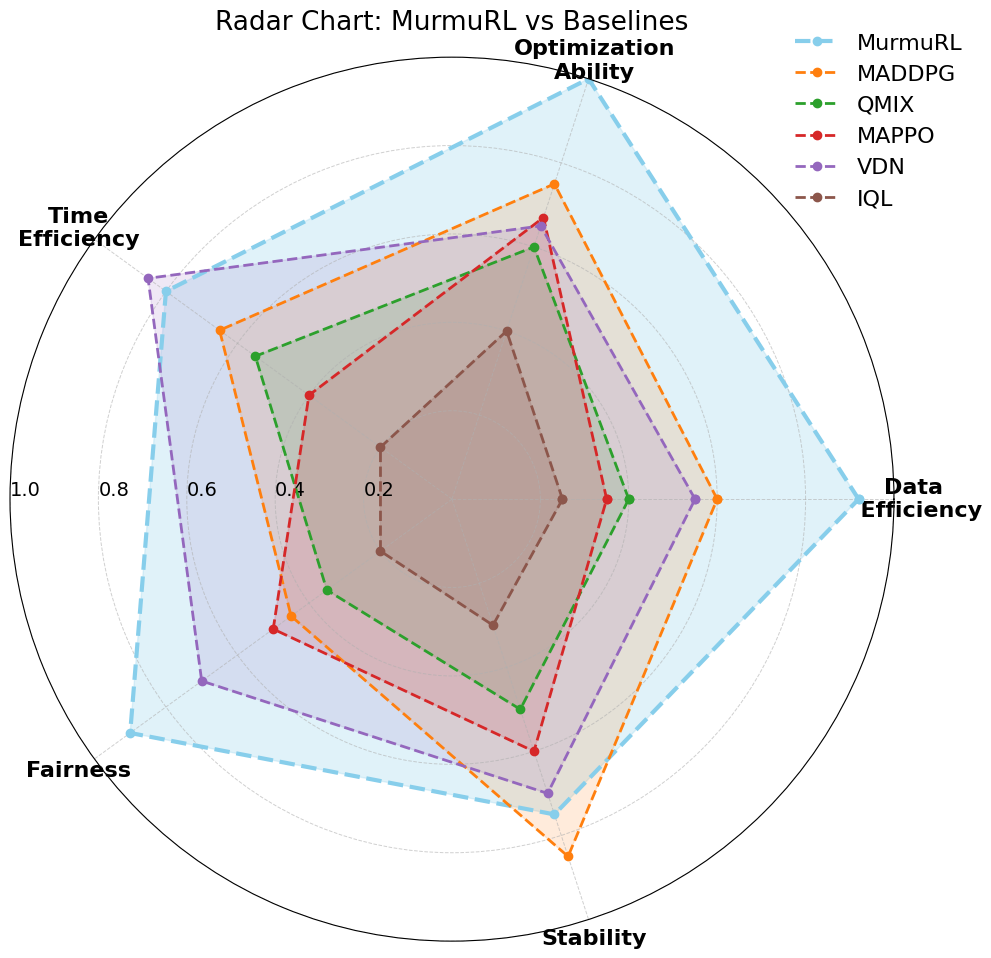}
  \caption{Radar chart comparing performance across five metrics. \texttt{MurmuRL} (blue) shows balanced superior performance compared to baselines.}
  \label{fig:radar}
  \vspace{-0.5cm}
  \end{figure}
  
  The radar chart in Figure~\ref{fig:radar} provides a comparison across all evaluation metrics for the full \texttt{MurmuRL} framework.\texttt{MurmuRL} demonstrates balanced performance across all five dimensions: data efficiency (0.92), computational efficiency (0.88), fairness in water allocation (0.90), stability (0.75) and optimization ability (0.80).
  
  MADDPG shows competitive performance in stability (0.82) but falls short in data efficiency (0.41) and optimization ability (0.38). Other methods show varying strengths but consistently underperform across multiple metrics. QMIX achieves moderate optimization ability (0.55) but struggles with data efficiency. MAPPO demonstrates balanced but mediocre performance across all metrics.
\begin{figure}[t]
\centering
\includegraphics[width=0.4\textwidth]{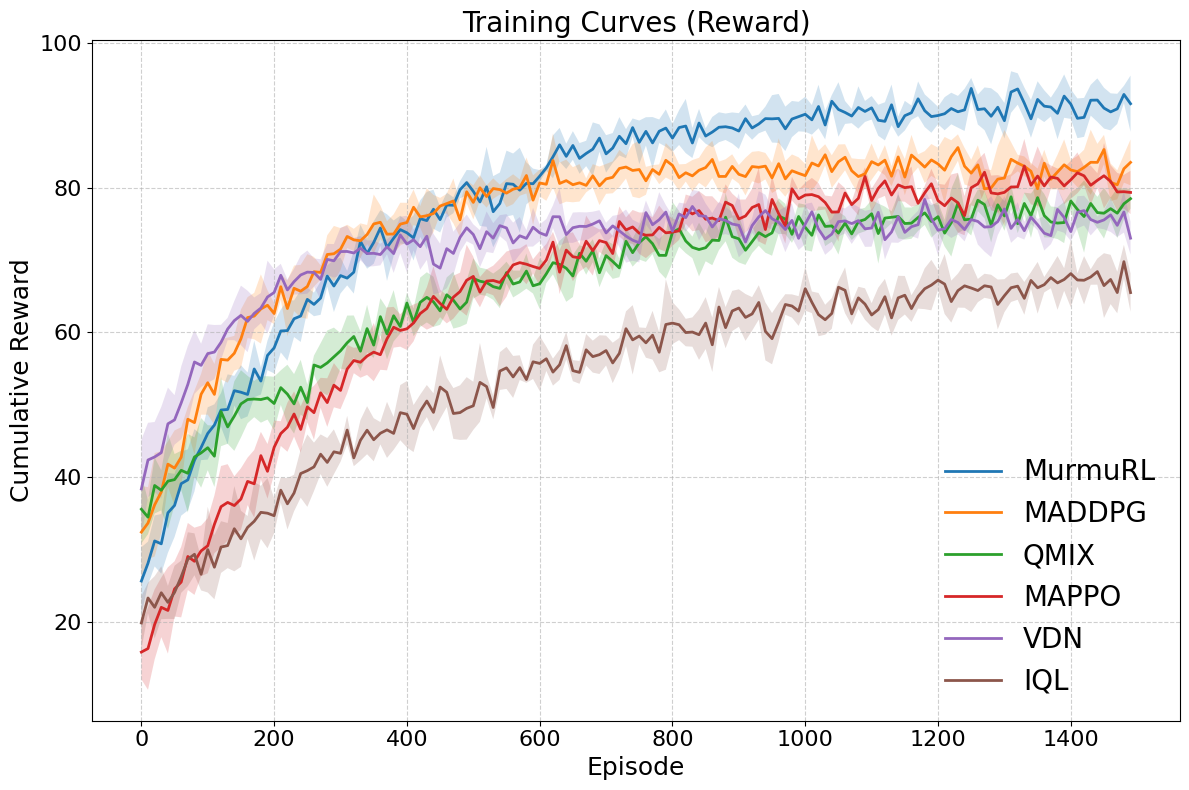}
\caption{Training curves showing cumulative rewards over 1500 episodes. \texttt{MurmuRL} (blue) achieves the highest final performance. MADDPG (orange) shows strong early learning but plateaus, while other methods demonstrate varying degrees of effectiveness.}
\label{fig:training_curve}
\vspace{-1.5em}
\end{figure}
Figure~\ref{fig:training_curve} illustrates the learning progression of different algorithms over 1500 training episodes. The \texttt{MurmuRL} approach shows a distinctive learning pattern. While MADDPG initially outperforms it during the first 400 episodes, \texttt{MurmuRL} continues to improve steadily throughout training. By approximately episode 600, it surpasses MADDPG and maintains its advantage until the end of training, ultimately achieving the highest cumulative reward of approximately 90.
This learning pattern aligns with our theoretical understanding of how emergent intelligence develops in murmuration-inspired systems. The initial phase requires the establishment of effective local interaction patterns, which takes time but creates a foundation for continued improvement. Once these patterns are established, the system's emergent properties allow it to discover increasingly effective strategies that centralized methods miss.

\begin{table}[t]
  \centering
  \resizebox{\linewidth}{!}{
    \begin{tabular}{lcccc}
    \toprule
    \textbf{Algorithm} & \makecell{\textbf{Training} \\ \textbf{Time (s/100 ep)}} & \makecell{\textbf{Sample} \\ \textbf{Efficiency}} & \makecell{\textbf{Resource} \\ \textbf{Utilization (\%)}} & \makecell{\textbf{Final} \\ \textbf{Reward}} \\
    \midrule
    \textbf{\texttt{MurmuRL}} & \textbf{77.44} & \textbf{319} & \textbf{73.21} & \textbf{89.25} \\
    MADDPG  & 55.01 & 253 & 94.63 & 82.03 \\
    QMIX    & 67.28 & 534 & 98.45 & 77.28 \\
    MAPPO   & 70.65 & 544 & 97.82 & 80.46 \\
    VDN     & 53.65 & 233 & 95.21 & 74.53 \\
    IQL     & 66.84 & 1354 & 90.36 & 67.42 \\
    \bottomrule
    \end{tabular}
  }
  \caption{Computational performance comparison}
  \label{tab:computation}
  \vspace{-1cm}
\end{table}

Table~\ref{tab:computation} presents the computational performance metrics for all methods. While \texttt{MurmuRL} requires somewhat more training time than some alternatives (77.44 compared to MADDPG's 55.01), it achieves substantially higher final rewards (89.25 vs. 82.03) and demonstrates excellent sample efficiency. \texttt{MurmuRL} reaches high performance after only 319 episodes, which is especially advantageous in real-world deployments where data collection can be expensive.

\subsection{Highlights}

Our experiments show several key advantages of \texttt{MurmuRL}:
\\\textbf{Super-linear Scaling:} As system size grows from 50×50 to 100×100 nodes, strategic diversity increases by 4.7× (compared to 4× growth in node count), demonstrating that complexity becomes an advantage rather than a limitation.
\\\textbf{Strategic Diversity with Stability:} \texttt{MurmuRL} maintains 4.3× higher strategy diversity than baseline approaches while preserving system-wide temporal stability.
\\\textbf{Balanced Performance:} \texttt{MurmuRL} achieves 8.8\% higher final performance while using 27\% less computing overhead compared to traditional approaches.

The most significant finding is that \texttt{MurmuRL} transforms what has traditionally been viewed as a weakness—system complexity—into a strength. Our 100×100 grid experiments provide compelling evidence: when the number of nodes increases by 4×, the clusters formed by our approach increase by 4.7×, demonstrating super-linear scaling of pattern complexity. This emergent property suggests that as water systems grow more complex, our approach's relative advantage over traditional methods grows correspondingly.

\section{Conclusion and Future Work}

\texttt{MurmuRL} integrates bio-inspired Alignment, Separation, and Cohesion rules with multi-agent reinforcement learning for decentralized reservoir management. Experimental results demonstrate that \texttt{MurmuRL} significantly outperforms conventional MARL techniques across key metrics with benefits that scale super-linearly as system complexity increases.

Our future research will extend in three directions: (1) validate \texttt{MurmuRL} on real-world hydrological data, (2) develop adaptive ASC weighting mechanisms for changing conditions, and (3) explore theoretical connections between murmuration dynamics and uncertainty robustness. By bridging Murmuration Intelligence and MARL systems, \texttt{MurmuRL} opens new pathways for managing systems in uncertain and changing scenarios.

\bibliographystyle{ACM-Reference-Format}
\bibliography{ref}

\end{document}